\documentclass{PoS}

\title{Charmonium physics with heavy ions:\\ experimental results}

\ShortTitle{Charmonium physics with heavy ions}

\author{\speaker{E. Scomparin}\\ 
        INFN, Torino (Italy)\\
        E-mail: \email{scomparin@to.infn.it}}

\abstract{Thirty years ago, the suppression of charmonium production in heavy-ion collisions was first proposed as an unambiguous signature for the formation of a Quark-Gluon Plasma. Since then, experiments at fixed-target accelerators (SPS) and hadronic colliders (RHIC, LHC) have investigated  this observable and discovered a wide range of effects, that have been related to the original proposal but at the same time have also prompted a strong development in the underlying theory concepts. In this contribution, I will review the main achievements of this field, with emphasis on recent results obtained by LHC experiments. } 

\FullConference{VIII International Workshop On Charm Physics\\
		5-9 September, 2016\\
		Bologna, Italy}

\begin{document}

\section{Introduction}

The production of charmonium states in heavy-ion collisions has attracted a lot of interest since the very beginning of studies related to the formation of a Quark-Gluon Plasma (QGP), a state of matter where quarks and gluons are no more confined inside hadrons. Due to the presence of free colour charges over a distance much larger than the typical hadron size, a screening of the colour forces between the c and $\overline{\rm c}$ quark was predicted to take place, leading to a suppression of charmonium states~\cite{Matsui:1986dk}. A $\sim$35\% suppression effect was first observed in 1986 by the NA38 Collaboration in \mbox{O-U} collisions at the SPS (at a center-of-mass energy per nucleon-nucleon collisions $\sqrt{s_{\rm NN}}=19.4$ GeV), by comparing the J/$\psi$ yields in central (head-on) and peripheral (grazing) collisions~\cite{Abreu:1988tp}. However, it was soon realized that alternative effects may lead to a similar observation. In particular, a break-up of the J/$\psi$ in the interaction with the dense hadronic gas, present after the system has cooled down below the critical temperature $T_{\rm c}\sim155$ MeV can also lead to an observable suppression~\cite{Armesto:1997sa}. In addition, effects related to the presence of a cold nuclear medium, both in the initial (nuclear shadowing, energy loss) and in the final state (break-up in cold nuclear matter) can also modify the J/$\psi$ yield~\cite{Gerschel:1988wn}.

In the following years, the production of J/$\psi$ and of the weakly bound $\psi(2S)$ states was studied in more detail at the SPS, making use of various nuclear beams (\mbox{Pb-Pb} and \mbox{In-In} collisions, by the NA50 and NA60 experiments, respectively~\cite{Alessandro:2004ap,Arnaldi:2007zz}) and also studying p-A collisions~\cite{Alessandro:2003pi}, in order to calibrate the size of cold nuclear matter (CNM) effects. At the end of this first round of experimental studies, it was clearly established that the suppression observed in nuclear collisions exceeded the effects that can be ascribed to CNM, and that the effects of the hot hadronic gas could hardly reproduce the measured suppression pattern as a function of the centrality of the collision.

Starting in 2000, with the availability of nuclear beams at the RHIC hadronic collider at much higher energies (\mbox{Au-Au} up to $\sqrt{s_{\rm NN}}=200$ GeV), a new era opened for quarkonium studies, and a further ``quantum jump'' happened in 2010, when \mbox{Pb-Pb} collisions were first studied at $\sqrt{s_{\rm NN}}=2.76$ TeV at the LHC. In the following Sections, I summarize some of the main achievements obtained at the RHIC and LHC colliders.

\section {Charmonium results at RHIC}

\begin{figure}[hbtp]
\begin{center}
\includegraphics[width=0.6\linewidth]{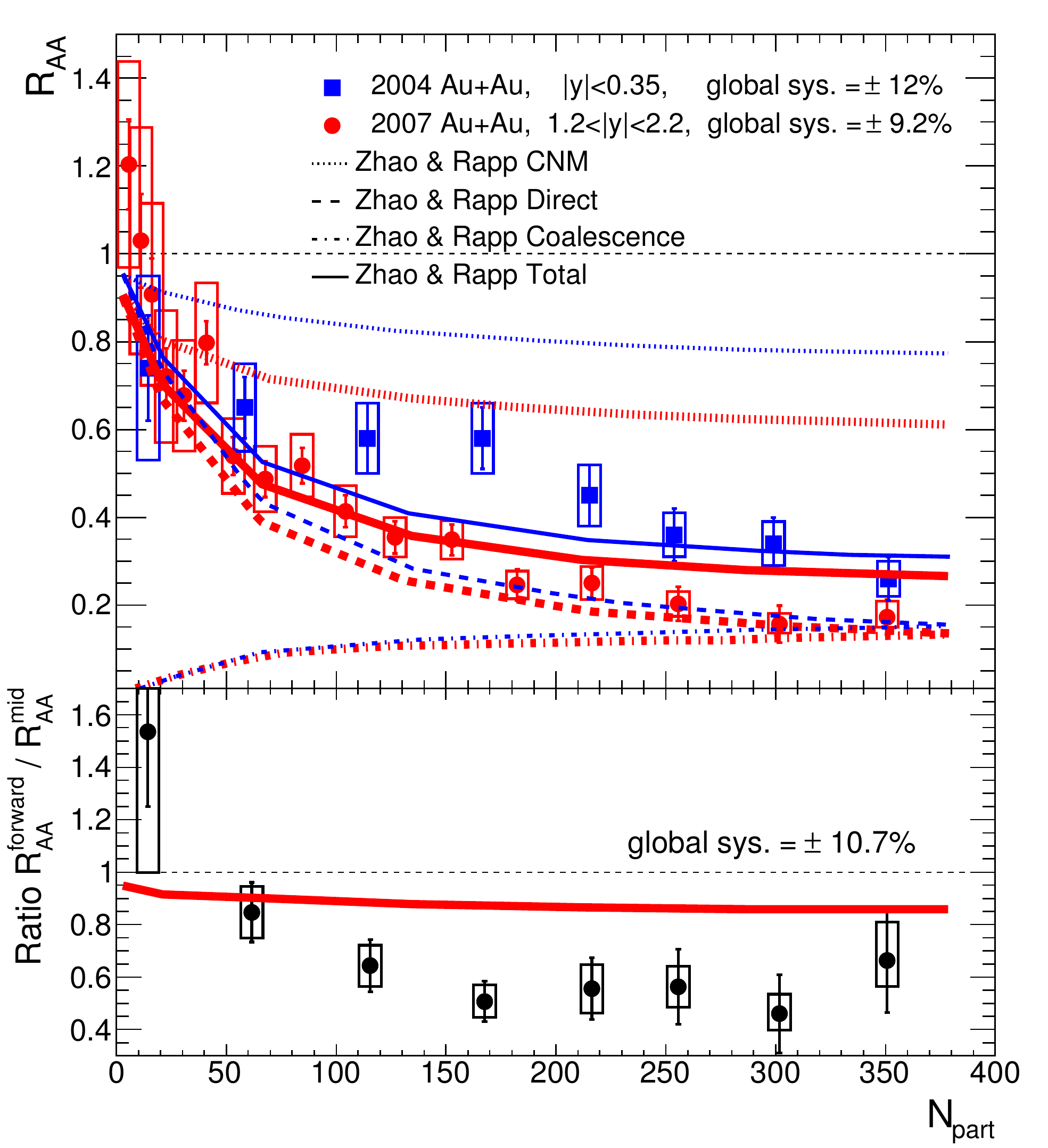} 
\caption{J/$\psi$ $R_{\rm AA}$ as a function of $N_{\rm part}$~\cite{Adare:2011yf}.
Model calculations by Zhao and Rapp are included for
both rapidity bins, incorporating cold and hot nuclear matter suppression as well as coalescence of c$\overline{\rm c}$
pairs.  The various line styles represent the different 
contributions to the total as laid out in the legend, while 
the two thicknesses represent the two rapidity ranges 
(thin blue is midrapidity and thick red is forward rapidity). 
The lower panel contains the ratio of forward rapidity to 
midrapidity for all points and curves in the upper panel.}
\label{fig:1}
\end{center}
\end{figure}

The PHENIX and STAR experiments have studied J/$\psi$ and $\psi(2S)$ production on a variety of collision systems, including (but not only) \mbox{Au-Au}, \mbox{Cu-Cu}, \mbox{d-Au} and, more recently, \mbox{U-U} and \mbox{p-A}. One of the early and most important results was the discovery of a strong rapidity ($y$) dependence of the J/$\psi$ suppression, with a significantly larger effect at forward $y$~\cite{Adare:2006ns}.
Such an observation is difficult to explain in a color-screening scenario, as the density of deconfined color charges should be larger at mid-rapidity and consequently lead to a stronger effect in that region. On the other hand, a natural explanation was proposed in terms of (re)combination of charm quarks, during the QGP phase and/or at phase boundary~\cite{BraunMunzinger:2000px,Thews:2000rj}. Such an effect is sizeable only when the charm quark density is relatively high, and is expected therefore to be more important at mid-rapidity, leading to a reduced suppression in that region. It becomes sizeable only at RHIC energies, since at the SPS the multiplicity of charm quarks is much smaller. Theoretical models that include a combination of suppression/(re)generation effects (Fig.~\ref{fig:1})  qualitatively describe the PHENIX results~\cite{Adare:2011yf}, shown in terms of the nuclear modification factor $R_{\rm AA}$ as a function of the number of  nucleons participating in the collision ($N_{\rm part}$). $R_{\rm AA}$ is the ratio between the yields in nucleus-nucleus and pp collisions, normalized to the number of binary nucleon-nucleon collisions ($N_{\rm coll}$). The latter quantity is calculated in a standard way, making use of the Glauber model. 

Since the bulk of c$\overline{\rm c}$ pairs is produced at low transverse momentum ($p_{\rm T}$), the (re)generation process is expected to enhance J/$\psi$ production in that transverse momentum region. The STAR Collaboration has shown the $p_{\rm T}$ dependence of $R_{\rm AA}$ for \mbox{Au-Au} collisions up to $p_{\rm T}>10$ GeV/$c$~(see~\cite{Adamczyk:2012ey} and more recent preliminary results). Remarkably, even for central events, no rise of the J/$\psi$ $R_{\rm AA}$ is seen in that region, an indication that (re)generation, although present, is probably not playing a dominant role at RHIC energy.

\begin{figure}[hbtp]
\begin{center}
\includegraphics[width=1.0\linewidth]{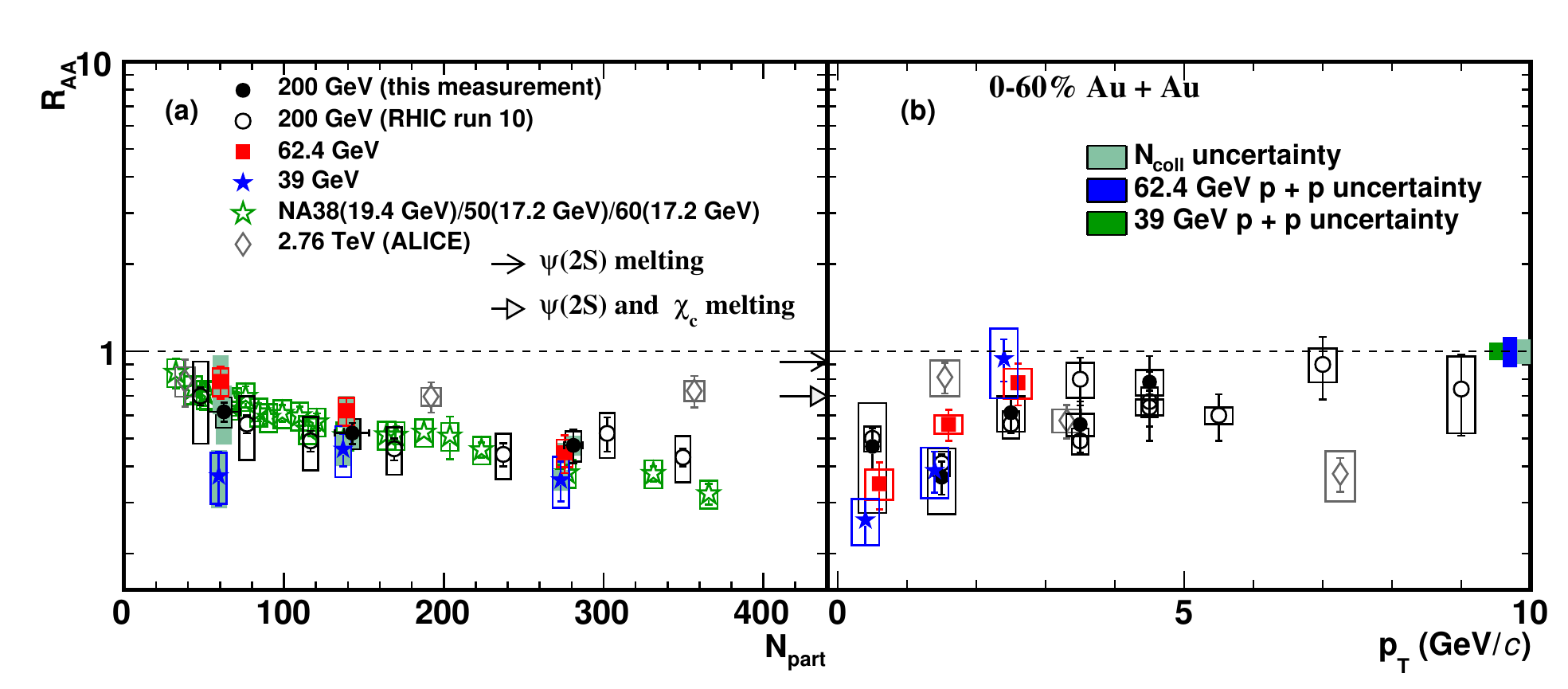} 
\caption{The results of J/$\psi$ $R_{\rm AA}$ as a function of
$N_{\rm part}$ (a) and $p_{\rm T}$ (b) in \mbox{Au-Au} collisions at
$\sqrt{s_{\rm NN}}$= 39, 62.4 and 200 GeV~\cite{Adamczyk:2016srz}. Statistical and systematic uncertainties are represented by error bars and boxes, respectively.  The shaded bands indicate the uncertainties from $N_{\rm coll}$ and the uncertainties for the derived baselines for $\sqrt{s_{\rm NN}}$=39 and 62.4 GeV. The ALICE points are also included. The expected ratio of feed-down J/$\psi$ from higher chamonium states to inclusive J/$\psi$ is also shown.}
\label{fig:2}
\end{center}
\end{figure}

It is interesting to note that comparing the suppression patterns, as a function of centrality and $p_{\rm T}$, for various center-of-mass energies, which span the range between top SPS and top RHIC energy (Fig.~\ref{fig:2}), the results are found to be qualitatively consistent~\cite{Adamczyk:2016srz}. This observation is possibly related to a balance of suppression effects, which are expected to increase with $\sqrt{s_{\rm NN}}$, and (re)generation, which also increases with $\sqrt{s_{\rm NN}}$, but acts in the opposite direction.

Clearly, CNM effects can play a sizeable role in the J/$\psi$ suppression observed at RHIC energies. A study of \mbox{d-Au} collisions at $\sqrt{s_{\rm NN}}$=200 GeV carried out by PHENIX~\cite{Adare:2012qf} has shown that $R_{\rm dAu}$ displays a similar behavior at mid and forward (d-going) rapidity, showing suppression at low $p_{\rm T}$ with a
gradual increase to a value consistent with 1. The $R_{\rm dAu}$ at backward (Au-going) rapidity has a different distribution with $p_{\rm T}$, showing a more rapid increase from suppression to a value of 1, and transitioning to
$R_{\rm dAu}>1$ above 2 GeV/$c$. Therefore, one can expect these CNM effects to play a role in determining the overall suppression observed in nucleus-nucleus collisions. However, no quantitative attempt to use the $R_{\rm dAu}$ results in such a way to determine the size of CNM effects in $R_{\rm AA}$ has been carried out for the moment.

Finally, the production of $\psi(2S)$ has been studied in \mbox{d-Au}~\cite{Adare:2013ezl} and, more recently, in \mbox{p-A} collisions~\cite{Adare:2016psx} (low statistics has prevented for the moment accurate studies in \mbox{Au-Au}). By comparing the $R_{\rm dAu}$ values for $\psi(2S)$ and J/$\psi$, a stronger suppression for $\psi(2S)$ is found for central \mbox{d-Au} collisions. Since initial state effects such as nuclear shadowing should approximately be of the same size for the two mesons, this difference has to be ascribed to final state effects. The influence of cold nuclear matter which the resonances cross should be negligible, since at RHIC energy the crossing time in CNM is much shorter than the resonance formation time. Consequently, this result shows that in \mbox{d-Au} collisions the particles produced in the interactions are responsible for the dissociation of the $\psi(2S)$. Due to the weakness of the $\psi(2S)$ binding, a hadronic medium (and therefore not necessarily a QGP) is probably sufficient to generate the observed effect.    

\section{Charmonium results at LHC}

\begin{figure}[hbtp]
\begin{center}
\includegraphics[width=0.48\linewidth]{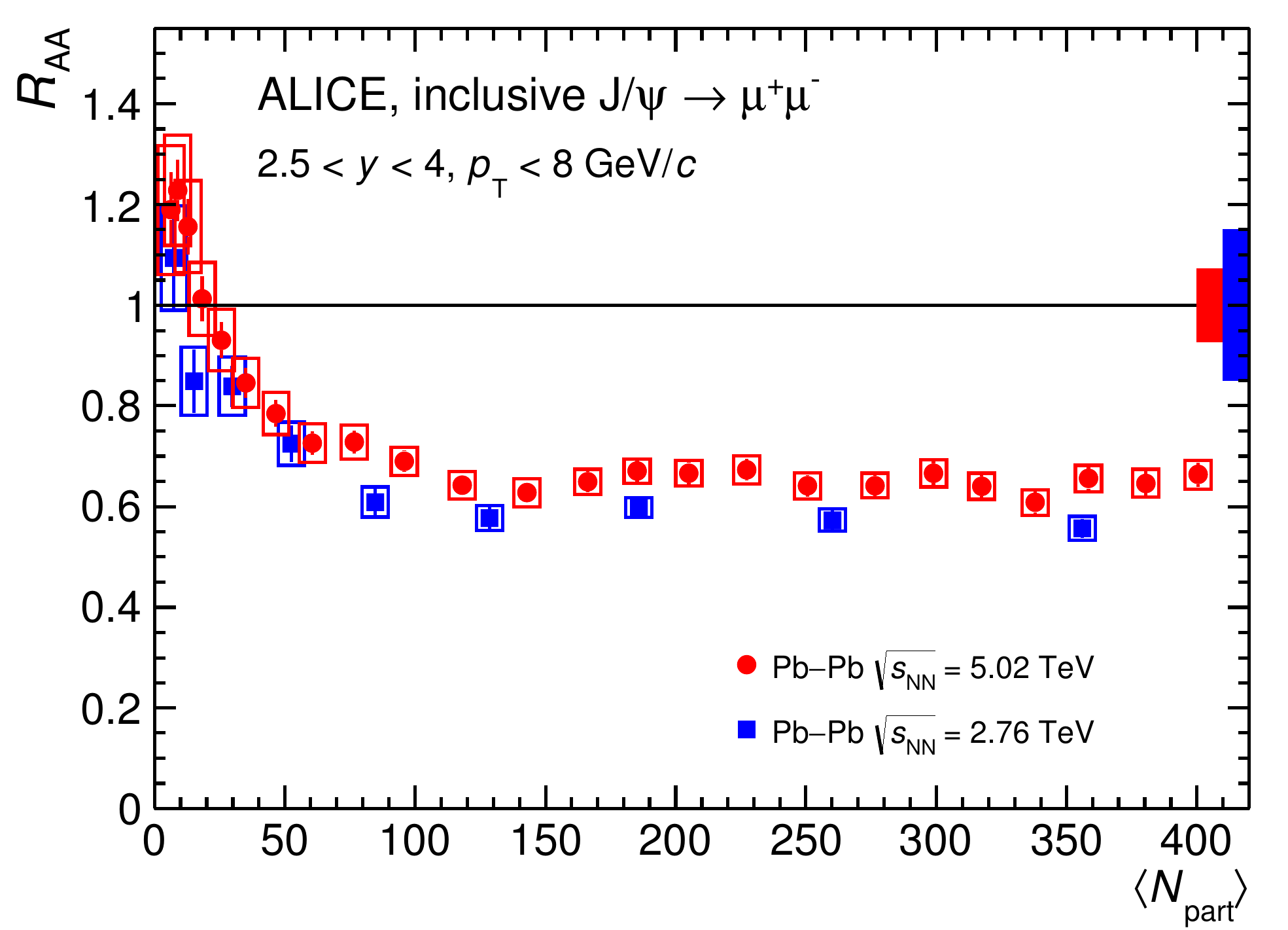} 
\includegraphics[width=0.48\linewidth]{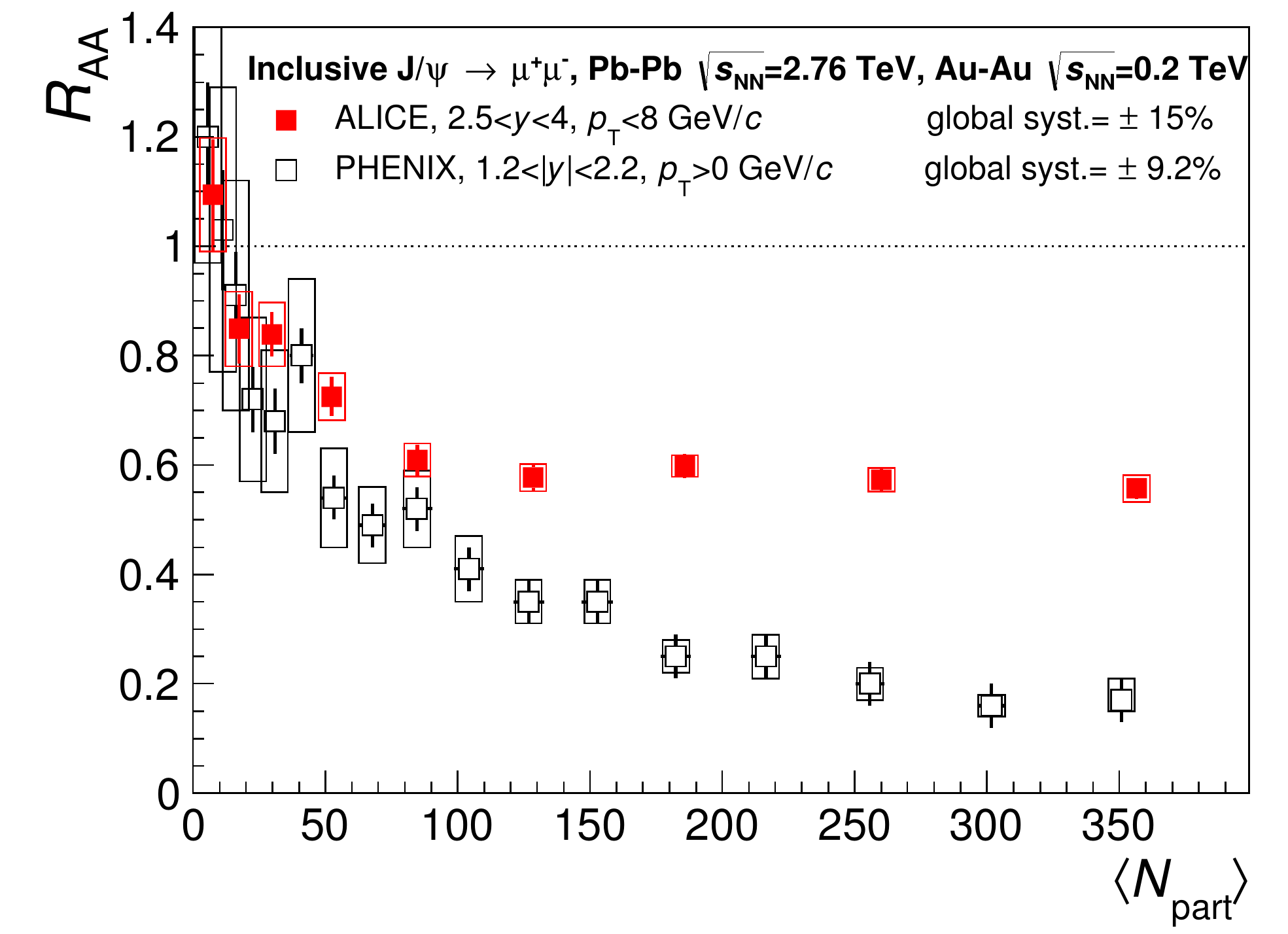} 
\caption{(Left) The nuclear modification factor for inclusive J/$\psi$ 
production, as a function of centrality, measured by ALICE in Pb-Pb collisions~\cite{Adam:2016rdg}. The error bars represent statistical uncertainties, the boxes around the points uncorrelated systematic uncertainties, while correlated global uncertainties are shown as a filled box around $R_{\rm AA}=1$; (right) Inclusive J/$\psi$ $R_{\rm AA}$ as a function of the number of participant nucleons measured in \mbox{Pb-Pb} collisions at $\sqrt{s_{\rm NN}}=2.76$ TeV~\cite{Abelev:2013ila}, compared to the PHENIX measurement in \mbox{Au-Au} collisions at $\sqrt{s_{\rm NN}}=0.2$ TeV.}
\label{fig:3}
\end{center}
\end{figure}

\begin{figure}[hbtp]
\begin{center}
\includegraphics[width=0.54\linewidth]{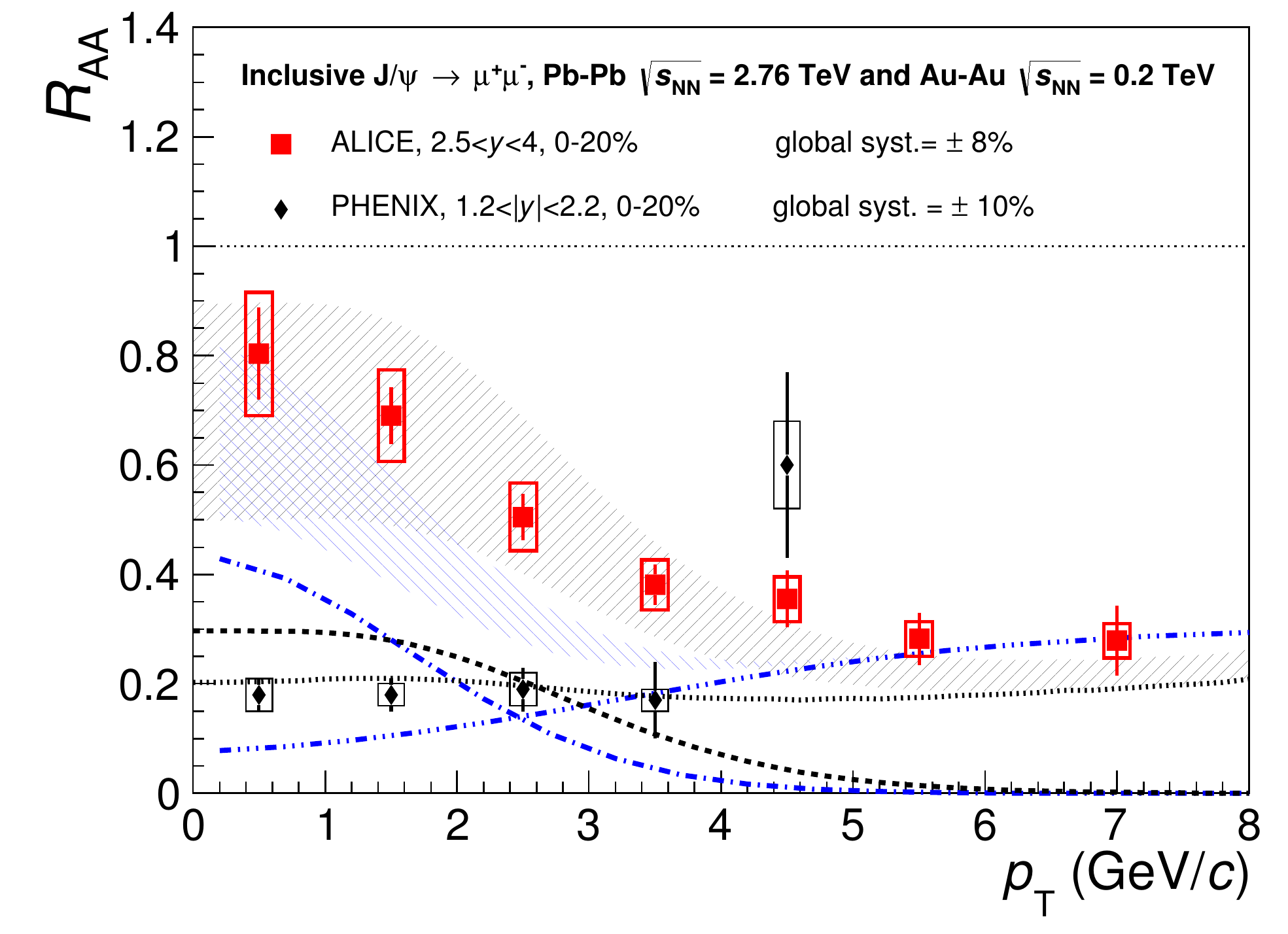} 
\includegraphics[width=0.44\linewidth]{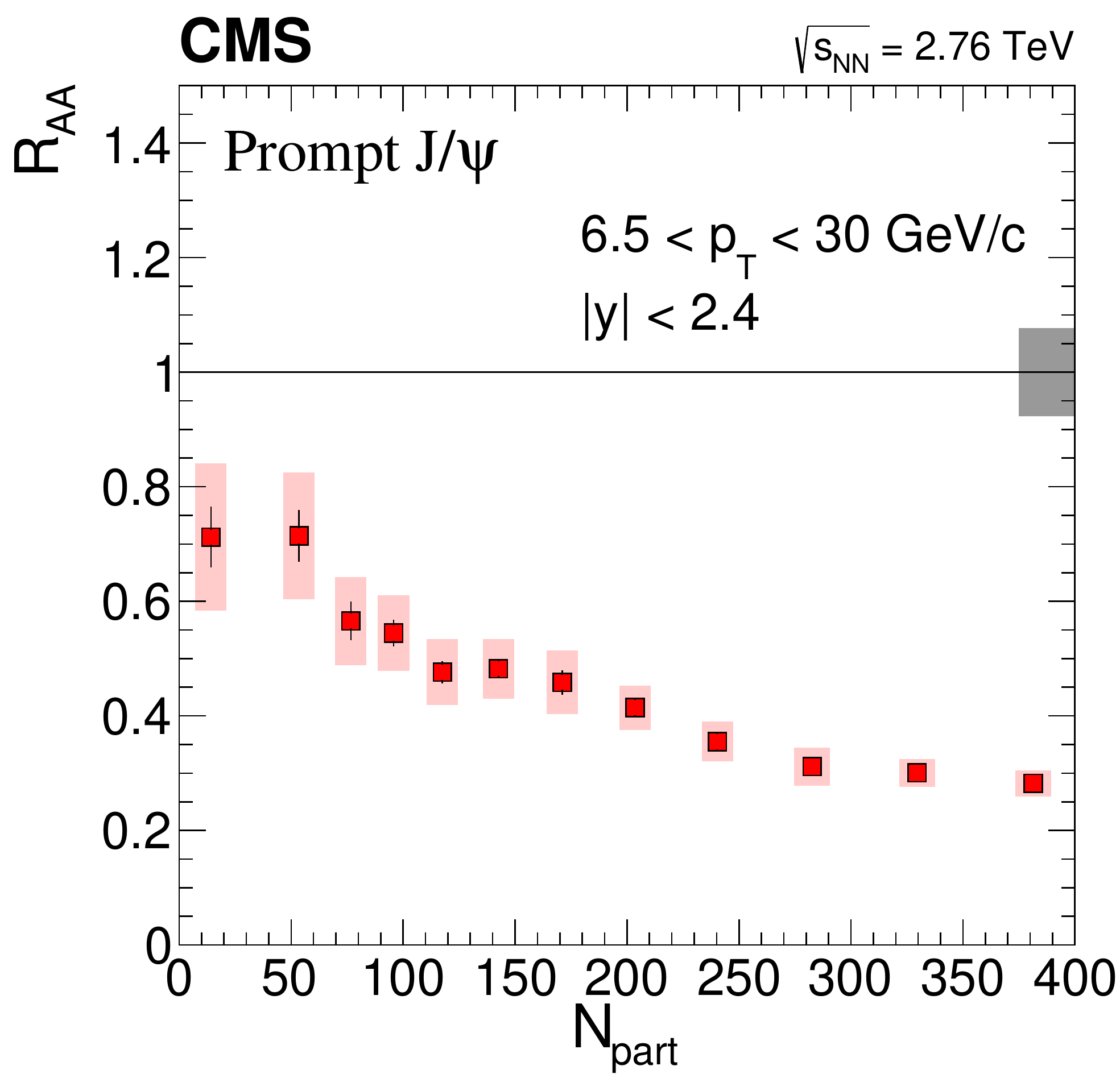} 
\caption{(Left) ALICE inclusive J/$\psi$ $R_{\rm AA}$ as a function of $p_{\rm T}$ for $2.5<y<4$ in the centrality class 0-20\%~\cite{Adam:2015isa} compared to transport models and to PHENIX results; (right) High-$p_{\rm T}$ prompt J/$\psi$ $R_{\rm AA}$ as a function of centrality, measured by CMS~\cite{Khachatryan:2016ypw}. The bars (boxes) represent statistical (systematic) point-by-point uncertainties. The gray boxes plotted on the right side at $R_{\rm AA}=1$ represent the scale of the global uncertainties.}
\label{fig:4}
\end{center}
\end{figure}

With the advent of the LHC, nucleus-nucleus (\mbox{Pb-Pb}) collisions at $\sqrt{s_{\rm NN}}=2.76$ TeV (run 1) and, more recently, $\sqrt{s_{\rm NN}}=5.02$ TeV (run 2) have become available, with a jump of more than one order of magnitude in collision energy with respect to RHIC. In Fig.~\ref{fig:3} (left),  results obtained by ALICE on the centrality dependence of the inclusive J/$\psi$  $R_{\rm AA}$ are shown~\cite{Adam:2016rdg}. The rapidity coverage is $2.5<y<4$, and the results, corresponding to $p_{\rm T}<8$ GeV/$c$, are dominated by low-$p_{\rm T}$ J/$\psi$. A suppression can be seen, increasing up to $N_{\rm part}\sim 100$ and then saturating. Within uncertainties, no dependence on $\sqrt{s_{\rm NN}}$ can be singled out. In Fig.~\ref{fig:3} (right) the $\sqrt{s_{\rm NN}}=2.76$ TeV results are compared with the corresponding forward-$y$ ($1.2<|y|<2.2$) PHENIX data~\cite{Abelev:2013ila}. The suppression observed at LHC is clearly smaller than at RHIC, an observation that points to a strong increase of (re)generation at LHC, overcoming the possible increase in the suppression related to the higher energy density expected at the LHC. The $p_{\rm T}$ dependence of $R_{\rm AA}$~\cite{Adam:2015isa}, compared in Fig.~\ref{fig:4} (left) to the corresponding PHENIX results and to theoretical calculations, shows a clear increase at low $p_{\rm T}$, which can be considered as a strong indication for a dominant contribution from J/$\psi$ (re)generation. This is confirmed by the model comparisons, where one can directly see the strong influence of the (re)generation process, which is maximum at low $p_{\rm T}$ and then quickly decreases.

The CMS experiment has also studied J/$\psi$ production in \mbox{Pb-Pb} collisions, exploring in particular the high-$p_{\rm T}$ region. In Fig.~\ref{fig:4} (right) the $R_{\rm AA}$ for prompt J/$\psi$ is shown, measured in the kinematic region $|y|<2.4$, $6.5<p_{\rm T}<30$ GeV/$c$. Contrary to the results shown in Fig.~\ref{fig:3}, dominated by low-$p_{\rm T}$ J/$\psi$, the suppression does not saturate up to central events, and it is much stronger than the one shown by ALICE. In addition, the $R_{\rm AA}$ values are systematically lower than those measured by STAR in a similar kinematic domain. Indeed, in such a high-$p_{\rm T}$ region, (re)generation effects become negligible and a stronger suppression may be expected at LHC energies, where the initial temperature of the system is higher and the  QGP lifetime is longer.

\begin{figure}[hbtp]
\begin{center}
\includegraphics[width=0.44\linewidth]{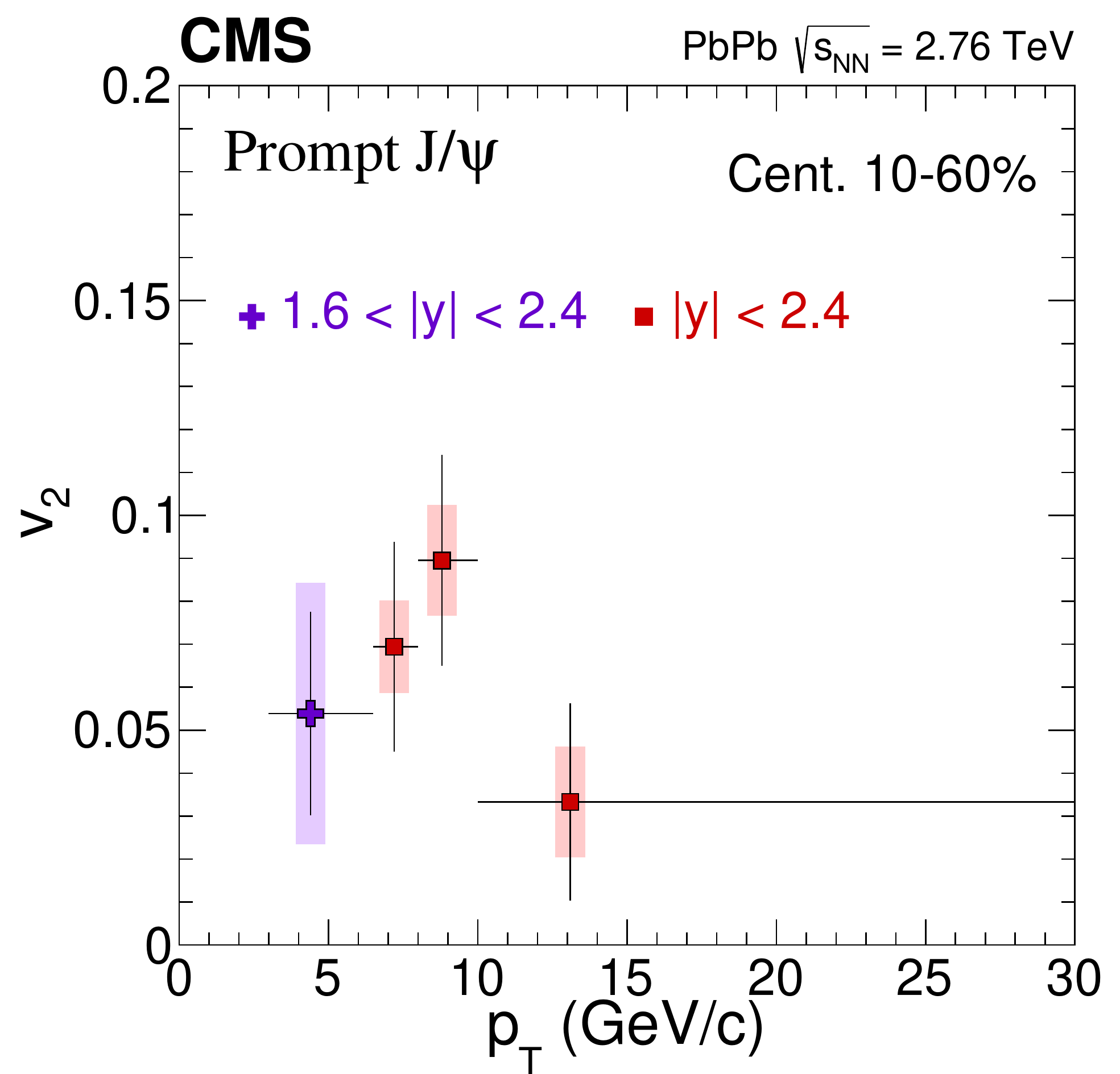} 
\includegraphics[width=0.54\linewidth]{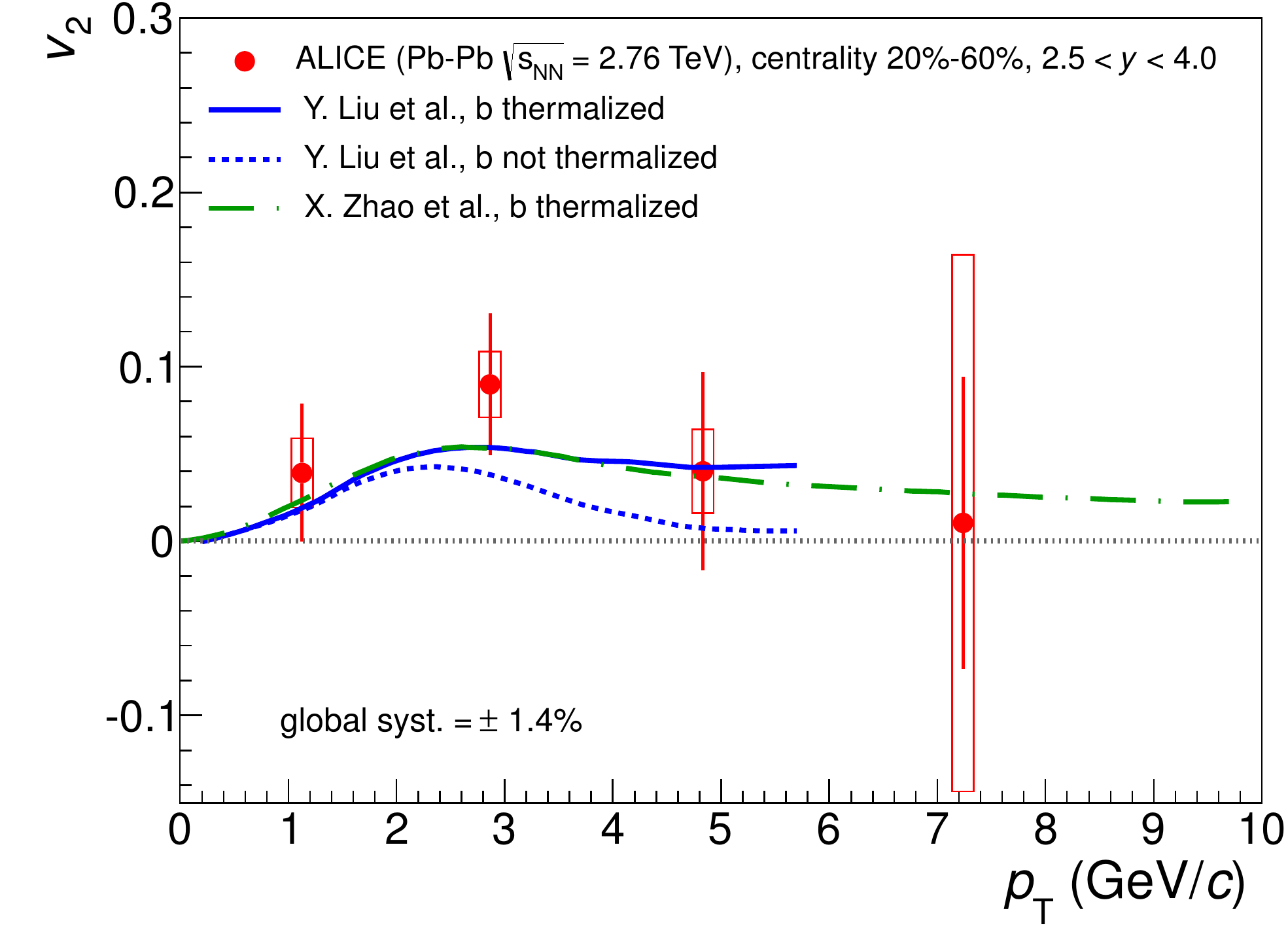} 
\caption{(Left) Prompt J/$\psi$ $v_{\rm 2}$ as a function of $p_{\rm T}$, measured by CMS~\cite{Khachatryan:2016ypw}. The bars (boxes) represent statistical (systematic) point-by-point uncertainties. Horizontal bars indicate the bin width; (right) ALICE results on inclusive J/$\psi$ $v_{\rm 2}$ vs $p_{\rm T}$ for non-central (20\%-60\%) \mbox{Pb-Pb} collisions at $\sqrt{s_{\rm NN}}=2.76$ TeV~\cite{ALICE:2013xna}, compared to calculations from two transport models.}
\label{fig:5}
\end{center}
\end{figure}

Another important observable in charmonium studies is the azimuthal anisotropy of the produced J/$\psi$ with respect to the reaction plane, expressed through the 2nd harmonics $v_{\rm 2}$ of the Fourier decomposition of the $\phi$-distribution. At low $p_{\rm T}$, a non-zero $v_{\rm 2}$ can be related to an indication for a thermalization of the recombining charm quarks, while at high $p_{\rm T}$ it is most likely connected with the different length of hot matter crossed by the c$\overline{\rm c}$ in and out of the reaction plane. A non-zero $v_{\rm 2}$ has indeed been observed by both ALICE (low $p_{\rm T}$)~\cite{ALICE:2013xna} and CMS (high $p_{\rm T}$)~\cite{Khachatryan:2016ypw} (see Fig.~\ref{fig:5}), even if more precise results are needed in order to extract a sharper physics message.

\begin{figure}[hbtp]
\begin{center}
\includegraphics[width=0.31\linewidth]{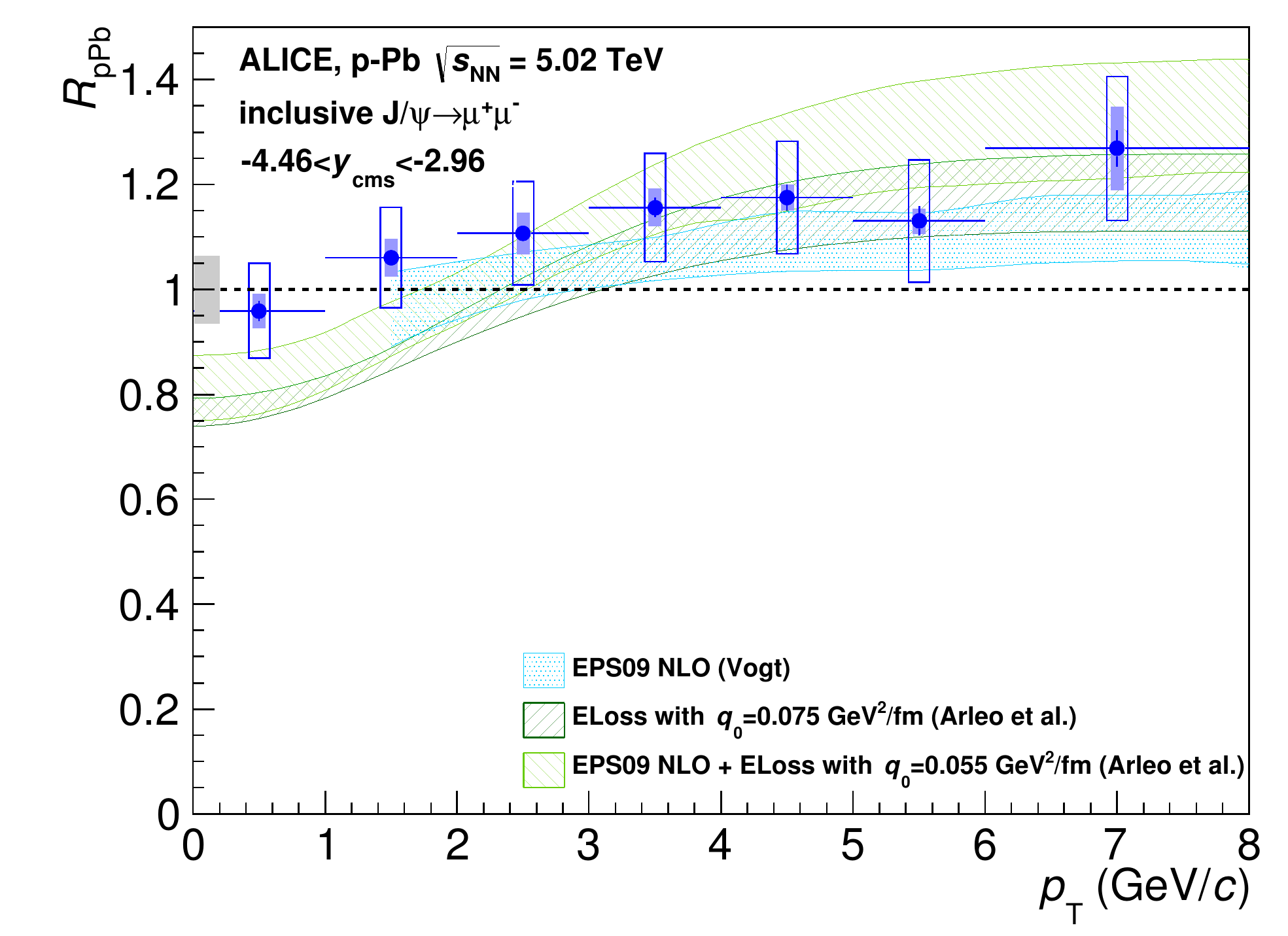} 
\includegraphics[width=0.31\linewidth]{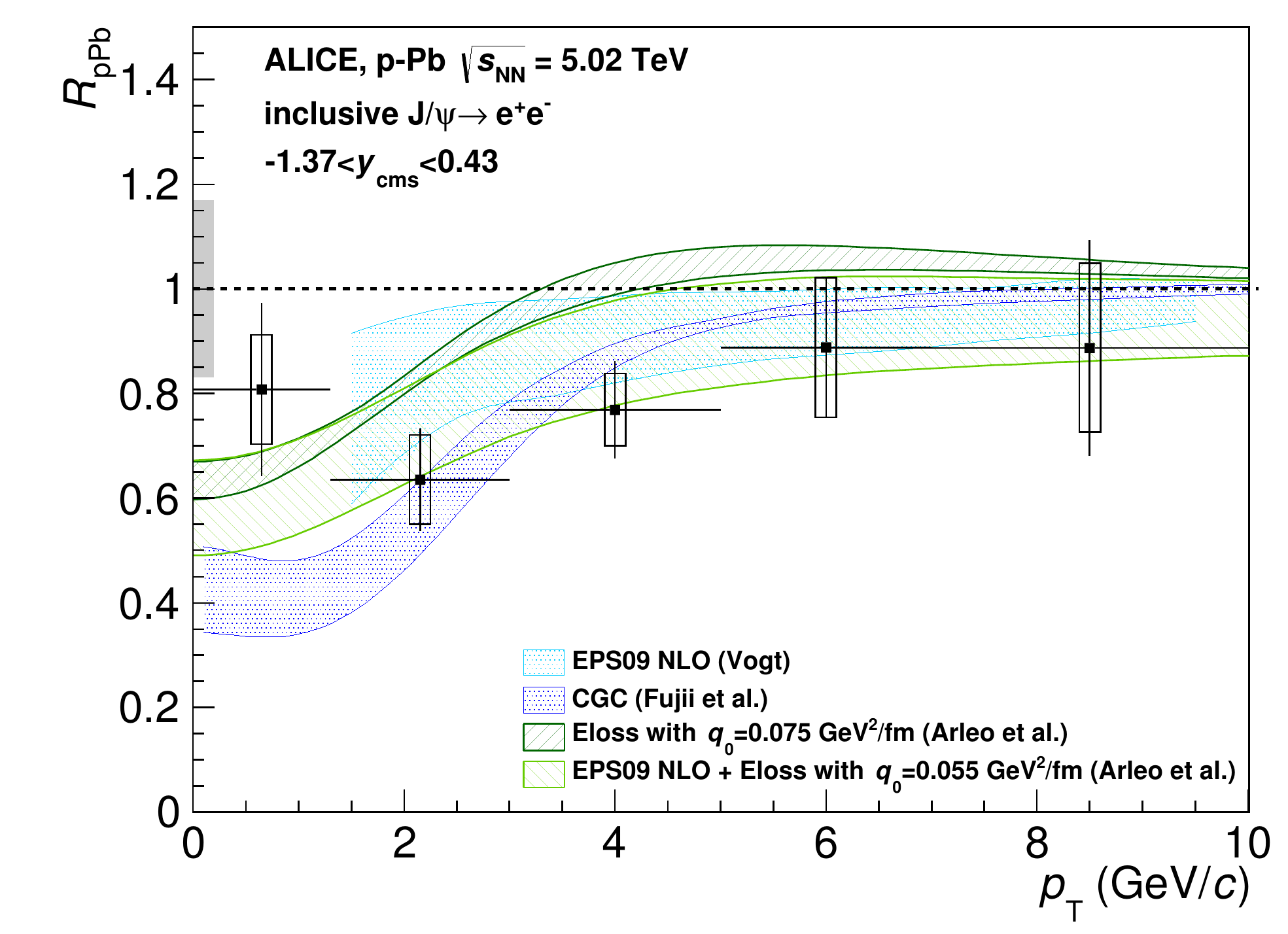} 
\includegraphics[width=0.31\linewidth]{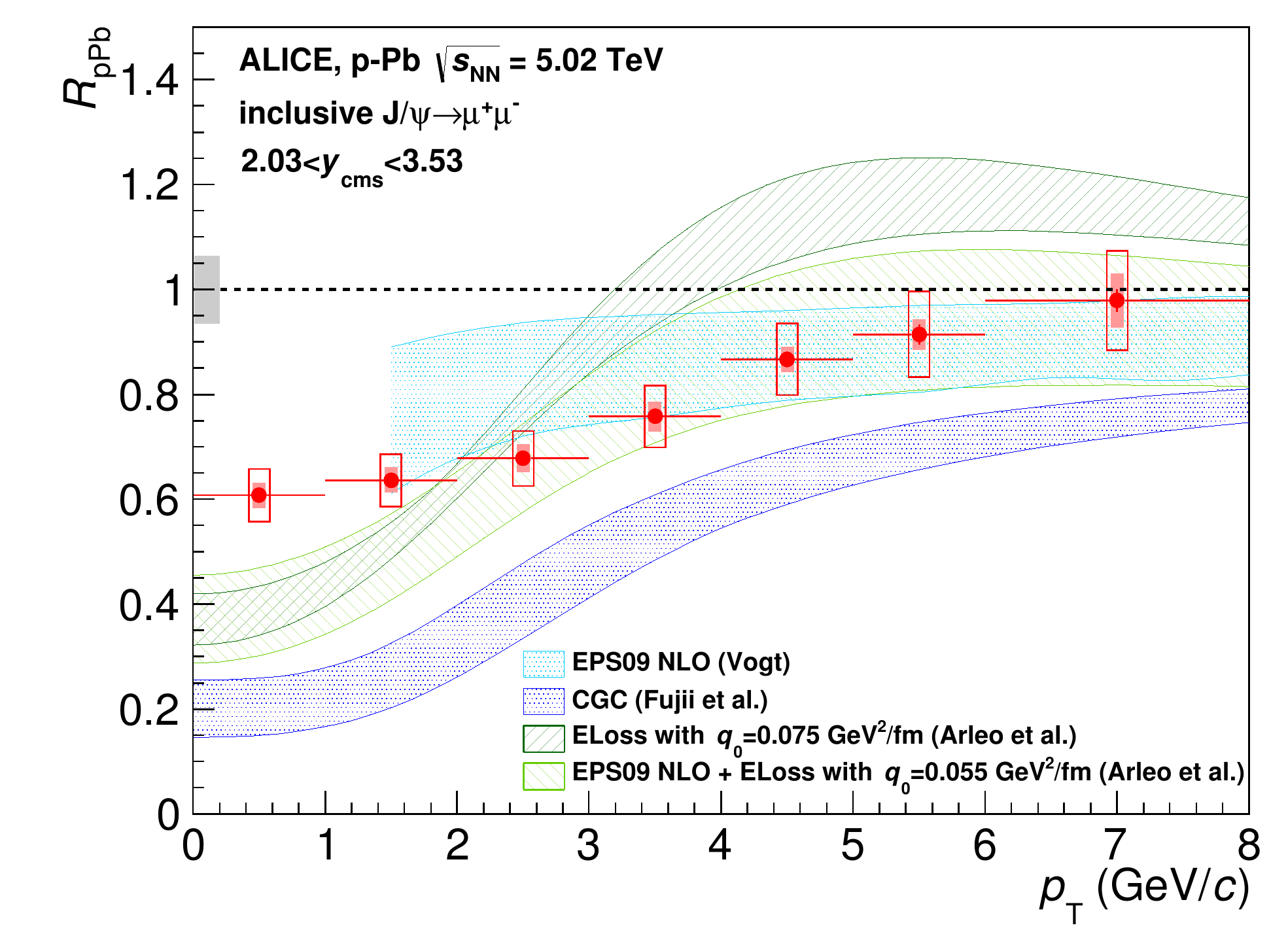} 
\caption{The J/$\psi$ nuclear modification factor as a function of $p_{\rm T}$ at backward (left), mid (center) and forward (right) rapidities~\cite{Adam:2015iga}, measured by ALICE. Statistical uncertainties are represented by vertical error bars, while open boxes correspond to uncorrelated uncertainties and the shaded areas to uncertainties partially correlated in $p_{\rm T}$. The boxes around $R_{\rm pPb}=1$ show the size of the correlated uncertainties. Results from various models are also shown.}
\label{fig:6}
\end{center}
\end{figure}

Also at LHC energies, CNM effects are expected to play a sizeable role. ALICE has studied the $p_{\rm T}$-dependence of $R_{\rm AA}$ in three $y$-ranges, corresponding to the p-going (forward), Pb-going (backward) and central rapidity regions. The results~\cite{Adam:2015iga} are shown in Fig.~\ref{fig:6} and indicate a rather strong suppression at low $p_{\rm T}$ at both forward and central rapidity, vanishing towards high $p_{\rm T}$, while at backward $y$  there might be an indication for a slight enhancement. Theoretical models can reproduce the results by means of nuclear shadowing effects and/or energy loss of the partons producing the J/$\psi$.   

\begin{figure}[hbtp]
\begin{center}
\includegraphics[width=0.7\linewidth]{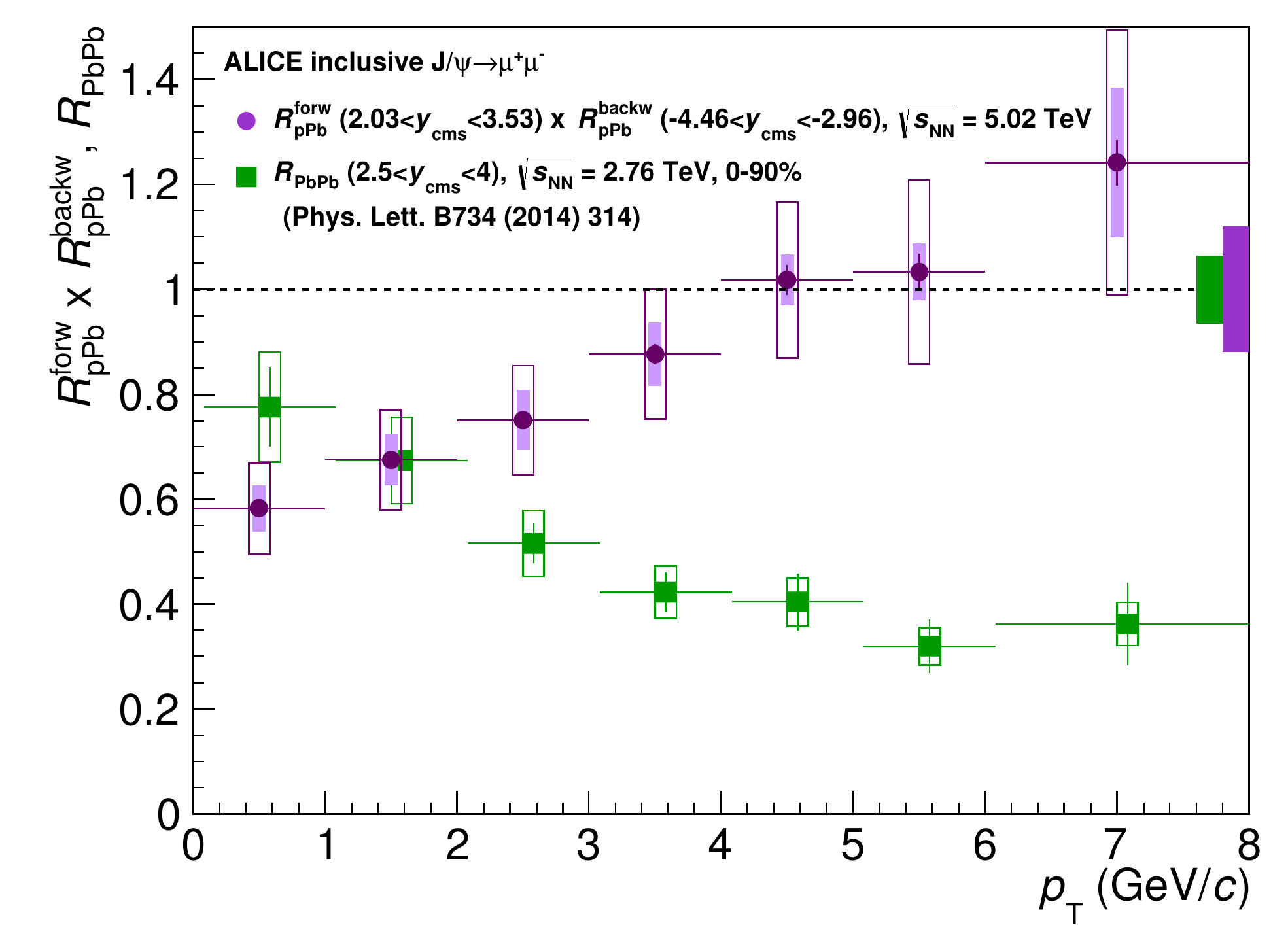} 
\caption{The estimate of the $p_{\rm T}$-dependence of CNM effects in \mbox{Pb-Pb} at forward rapidity~\cite{Adam:2015iga}, from ALICE data. The quantities are compared to $R_{\rm AA}$ measured in \mbox{Pb-Pb} collisions in the (approximately) corresponding y-ranges.}
\label{fig:7}
\end{center}
\end{figure}

The results of Fig.~\ref{fig:6} imply that CNM effects can influence the overall suppression observed in \mbox{Pb-Pb} collisions. Assuming that at LHC energy the \mbox{p-Pb} data can be described via a combination of nuclear shadowing and energy loss, a qualitative estimate of the size of CNM effects in \mbox{Pb-Pb} collisions at forward-$y$ can be obtained by forming the product of the $R_{\rm pPb}$ measured at forward and backward rapidity. The result is shown in Fig.~\ref{fig:7}, where such a product is compared  with the measured $R_{\rm AA}$, as a function of $p_{\rm T}$~\cite{Adam:2015iga}. One clearly sees that at high $p_{\rm T}$ CNM effects are expected to give a negligible contribution to the oberved strong J/$\psi$ suppression. On the contrary, at low $p_{\rm T}$ the observed suppression in \mbox{Pb-Pb} is of the same order of magnitude of that expected from CNM effects alone, corrsponding to a situation where suppression and (re)generation effects in \mbox{Pb-Pb} are roughly of the same size.

Results on the $\psi(2S)$ meson are also available at LHC energies. CMS has measured the so-called ``double ratio'' of $\psi(2S)$ to J/$\psi$ cross sections between \mbox{PbPb} and pp at $\sqrt{s_{\rm NN}}=2.76$ and 5.02 TeV~\cite{Khachatryan:2014bva,Sirunyan:2016znt}. The results show a suppression of the $\psi(2S)$ with respect to the J/$\psi$ at all centralities in the region $6.5<p_{\rm T}<30$ GeV/$c$, which can be expected due to the much weaker binding energy of the 2S state. When the low-$p_{\rm T}$ limit is pushed down to 3 GeV/$c$ an enhancement of the $\psi(2S)$ is visible for central events, but only at $\sqrt{s_{\rm NN}}=2.76$ TeV. On the contrary, $\sqrt{s_{\rm NN}}=5.02$ TeV results do not show such an enhancement. The rise of $R_{\rm AA}$ for central events in the intermediate $p_{\rm T}$ region has been interpreted in terms of (re)generation of $\psi(2S)$ occurring at a later time in the collision history with respect to J/$\psi$, in a moment when the collective expansion of the system has increased the average $p_{\rm T}$ of the flowing particles~\cite{Du:2015wha}. However, it is not straightforward to understand why the effect is seen only at the lower collision energy.

\begin{figure}[hbtp]
\begin{center}
\includegraphics[width=0.48\linewidth]{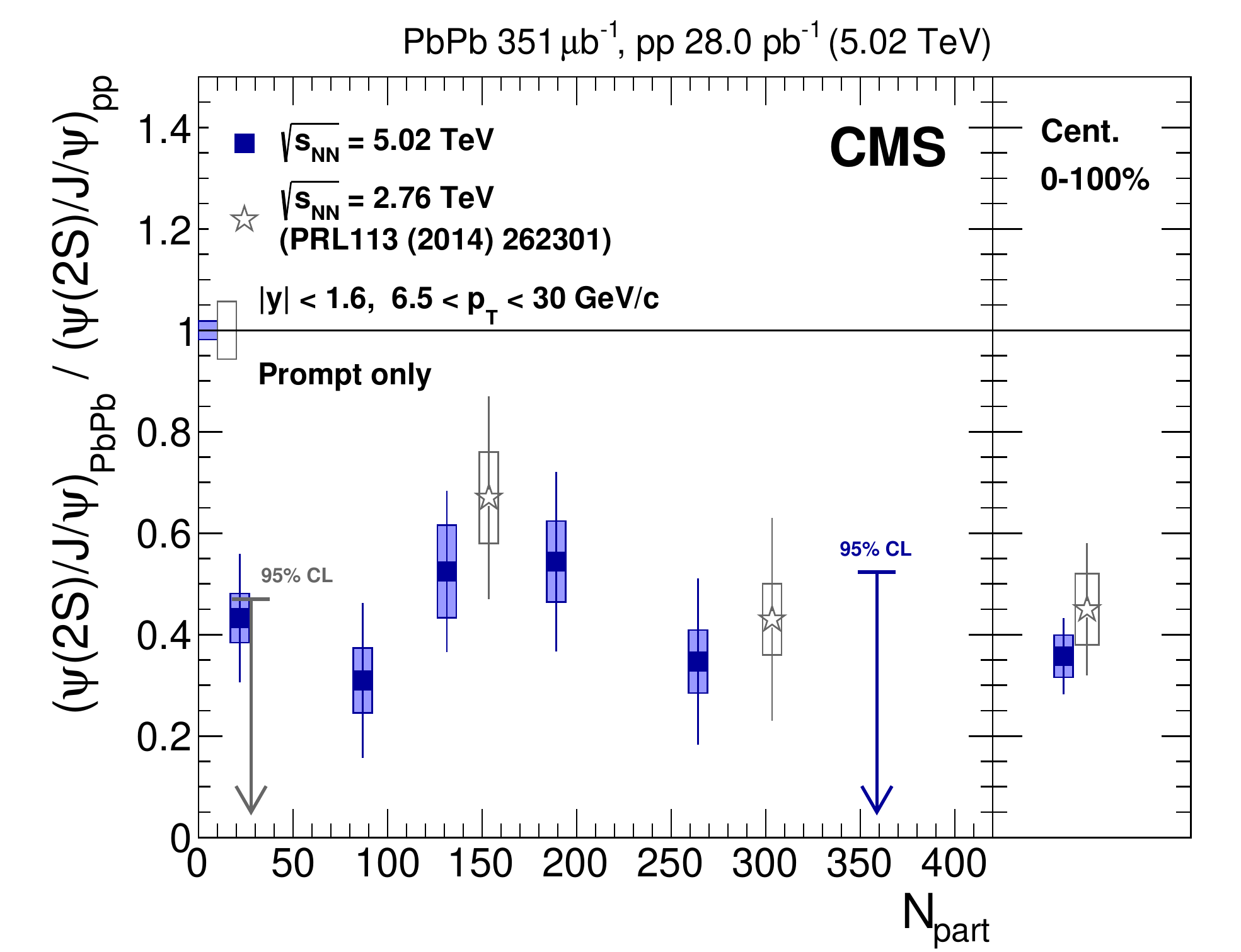} 
\includegraphics[width=0.48\linewidth]{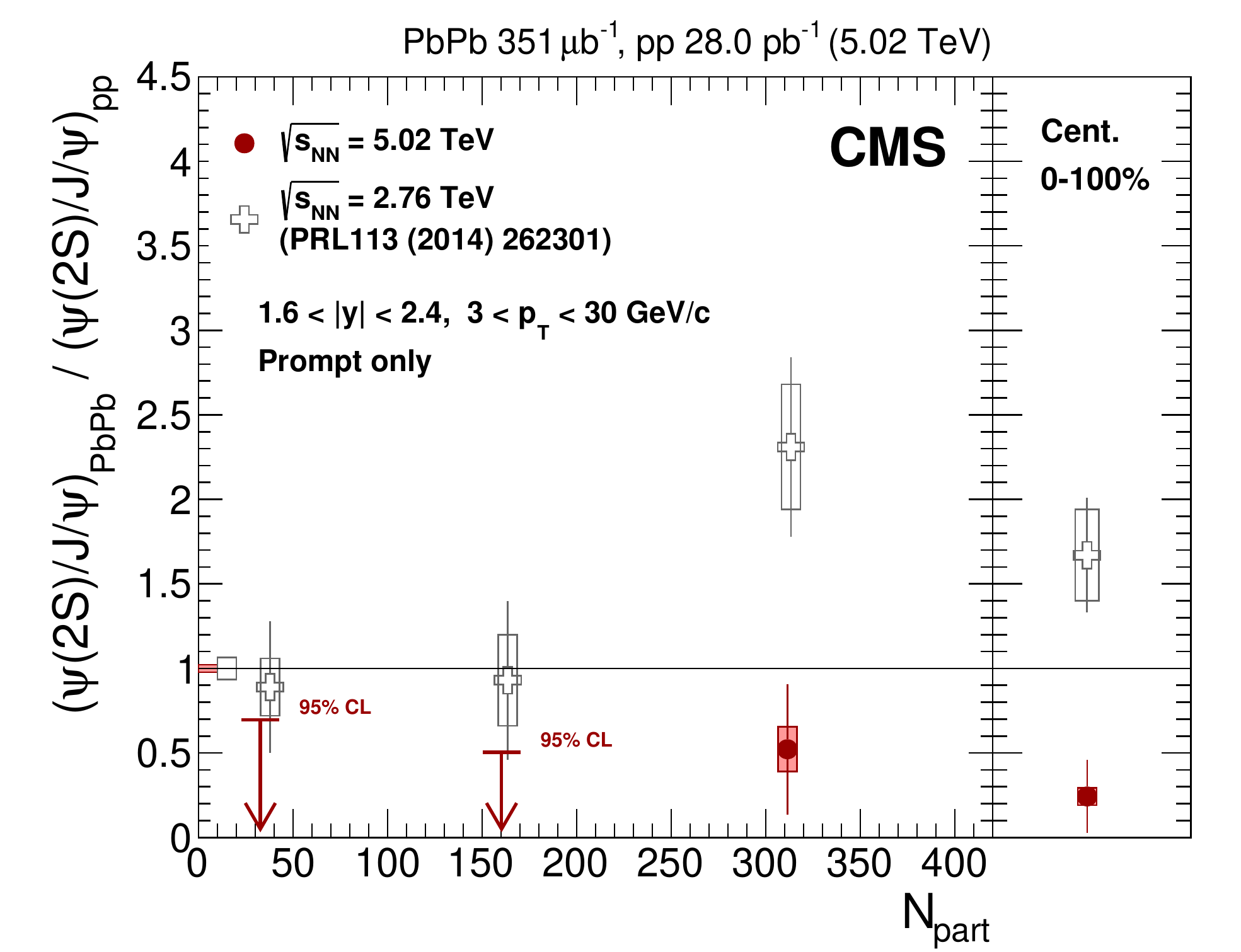} 
\caption{Event centrality dependence of $(N_{\psi(2S)}/N_{J/\psi})_{\rm PbPb}/(N_{\psi(2S)}/N_{J/\psi})_{\rm pp}$, for mid rapidity and in two $p_{\rm T}$ ranges~\cite{Khachatryan:2014bva,Sirunyan:2016znt}. Integrated values for all centrality are given in the right panels. The arrows represent 95\% CL upper limits in the bins where the measurement is consistent with 0. The vertical lines (bands) represent the statistical (systematic) uncertainties. The statistical and systematic uncertainties in the pp measurements, common to all points, are represented as boxes at unity.}
\label{fig:8}
\end{center}
\end{figure}

\begin{figure}[hbtp]
\begin{center}
\includegraphics[width=0.48\linewidth]{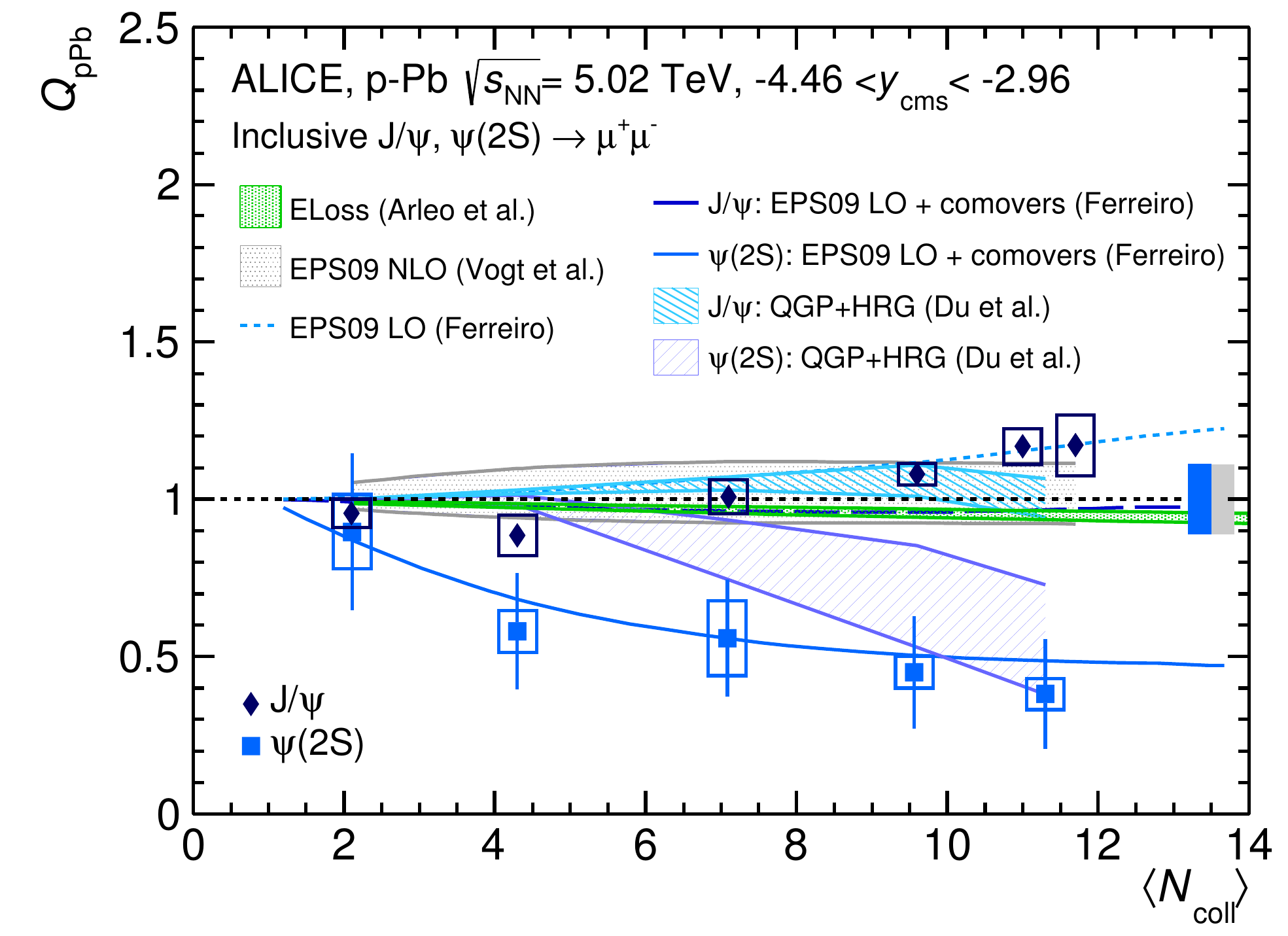} 
\includegraphics[width=0.48\linewidth]{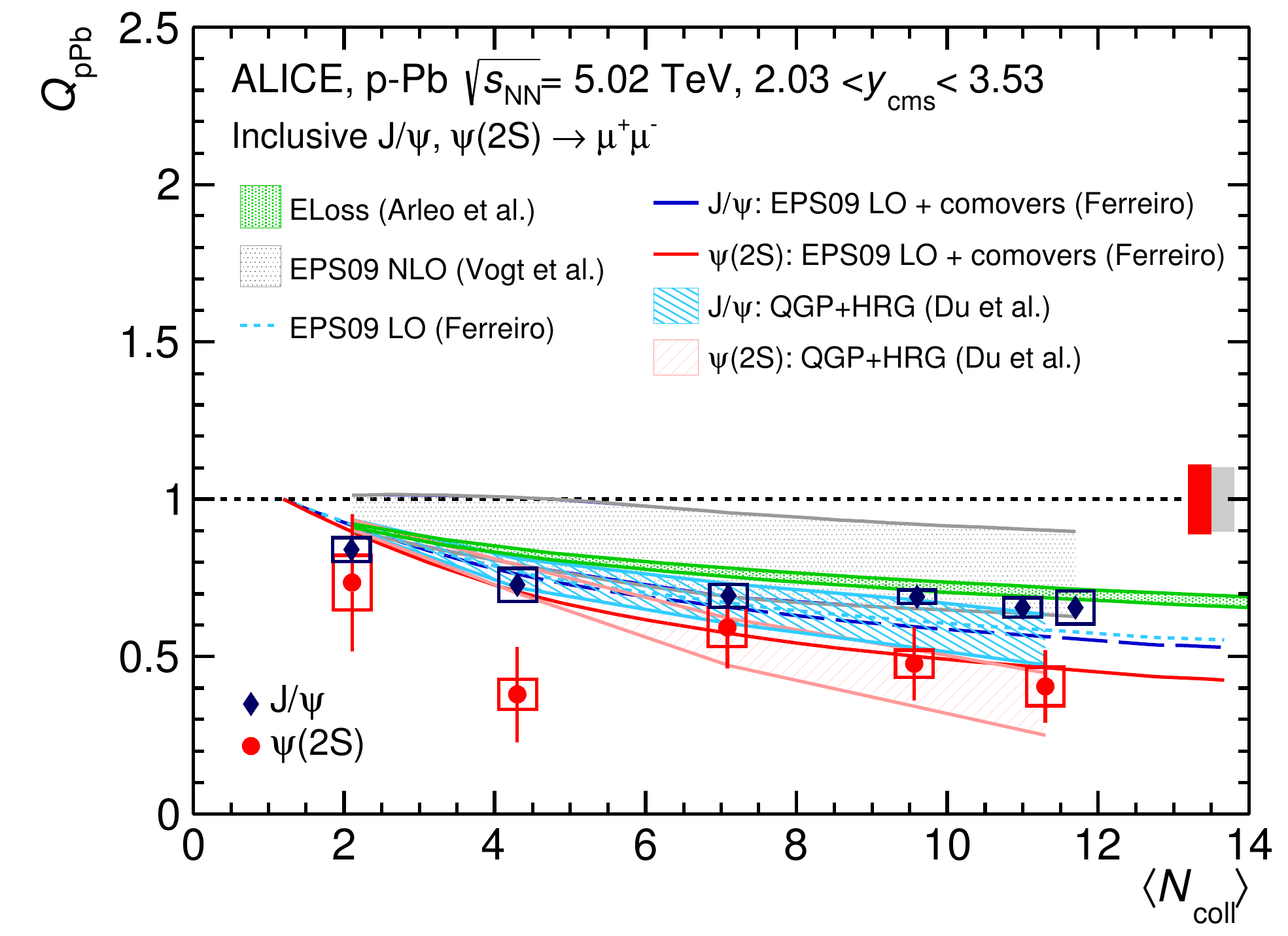} 
\caption{J/$\psi$ and $\psi(2S)$ nuclear modification factors, $Q_{\rm pPb}$, shown as a function of $N_{\rm coll}$ for the backward (left) and forward (right) rapidity regions and compared to theoretical models~\cite{Adam:2016ohd}. The boxes around unity correspond to the global $\psi(2S)$ systematic uncertainties at forward (red box) and backward (blue box) rapidities The grey box is a global systematic uncertainty common to both J/$\psi$ and $\psi(2S)$.}
\label{fig:9}
\end{center}
\end{figure}

The $\psi(2S)$ production has also been studied by ALICE in \mbox{p-Pb} collisions, in the forward and backward $y$ regions. In Fig.~\ref{fig:9} a comparison of the J/$\psi$ and $\psi(2S)$ nuclear modification factors is shown~\cite{Adam:2016ohd}, here indicated with the non-standard notation $Q_{\rm pPb}$. While in the p-going direction the $\psi(2S)$ only exhibits a slightly stronger suppression, in the Pb-going direction the effect is much stronger, and increases with the centrality of the \mbox{p-Pb} collision, here expressed through $N_{\rm coll}$. The effect is similar, but more evident, to the one observed by PHENIX in \mbox{d-Au} collisions, and represents a strong evidence for the presence of a dense system of particles, presumably of hadronic nature, which is able to dissociate the $\psi(2S)$ but not the strongly bound J/$\psi$. The suppression of the latter in the forward region has to be ascribed, as discussed above, to initial state effects as shadowing and energy loss.

\section {Bottomonium studies}

\begin{figure}[hbtp]
\begin{center}
\includegraphics[width=0.51\linewidth]{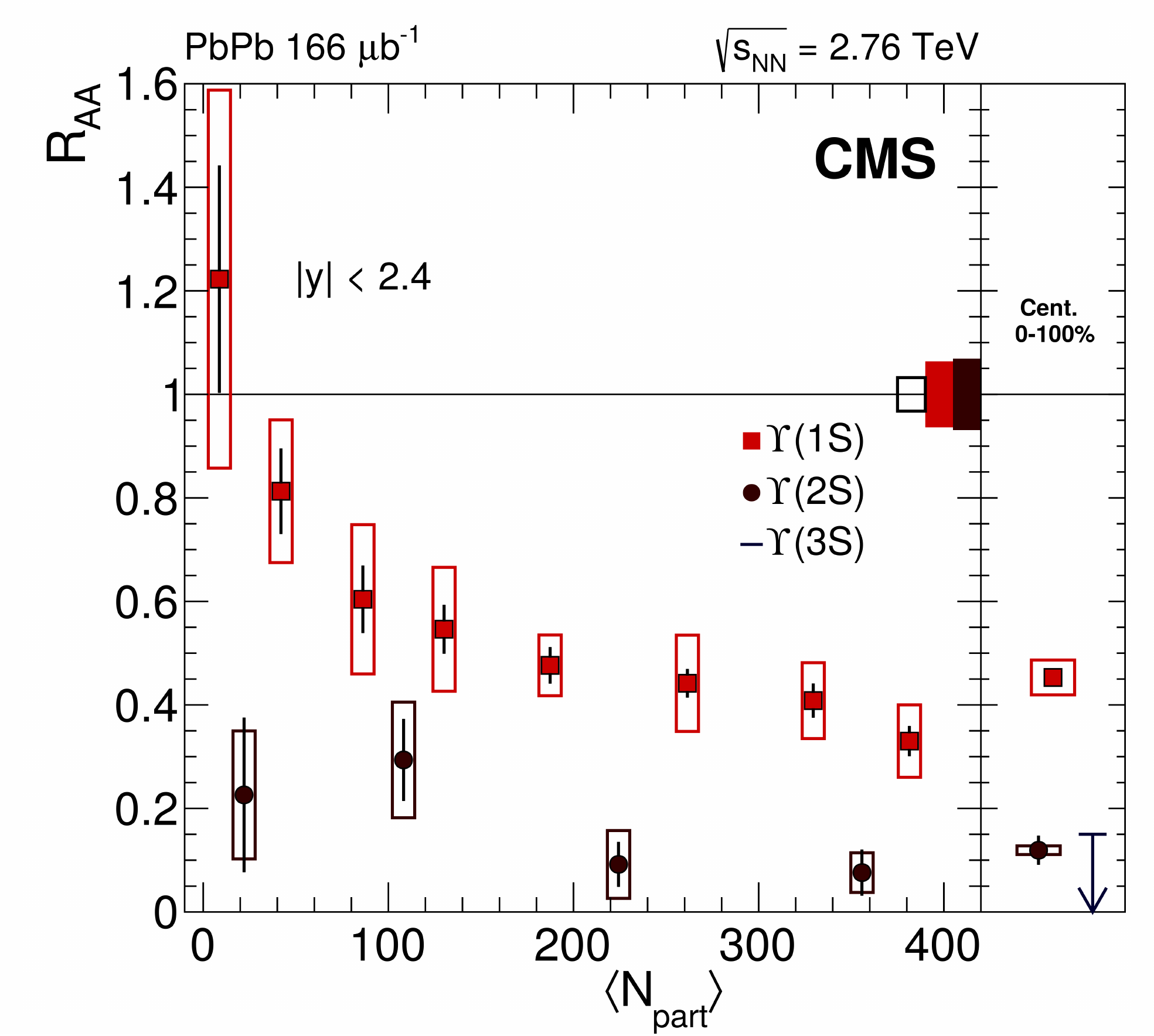} 
\includegraphics[width=0.46\linewidth]{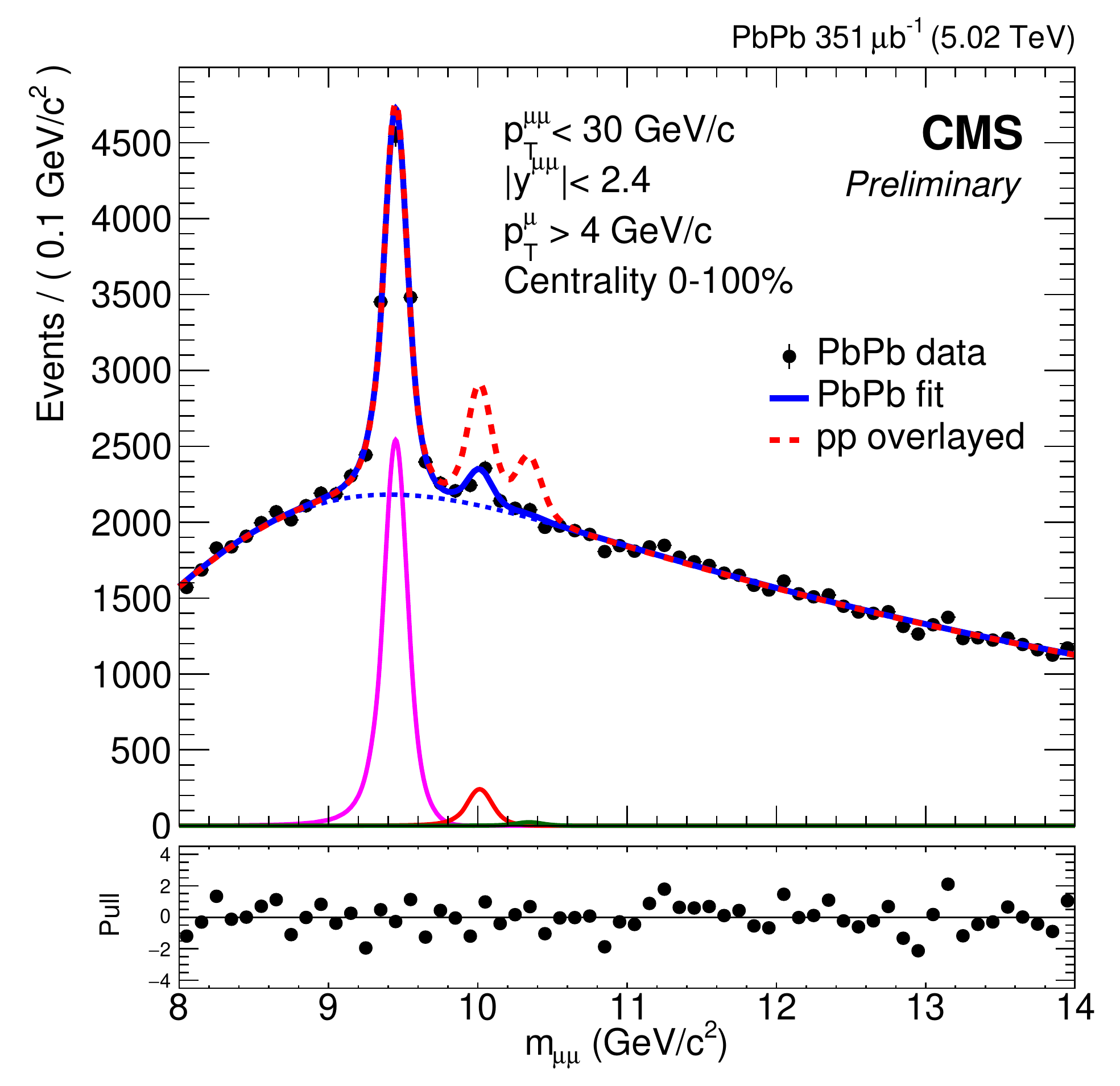} 
\caption{(Left) Nuclear modification factor of $\Upsilon(1S)$ and $\Upsilon(2S)$, measured by CMS in \mbox{Pb-Pb} collisions as a function of centrality, displayed as the number of participating nucleons~\cite{Khachatryan:2016xxp}. Statistical (systematic) uncertainties are displayed as error bars (boxes), while the global fully-correlated uncertainty is displayed as a box at unity; (right) Dimuon invariant mass distributions (black circles) for centrality-integrated \mbox{Pb-Pb} data, measured by CMS at $\sqrt{s_{\rm NN}}=5.02$ TeV~\cite{CMS:2016ayg}. The fit is shown as a solid blue line and the background component as a dashed blue line. $\Upsilon(1S)$, $\Upsilon(2S)$, and $\Upsilon(3S)$ are also depicted as solid magenta, red and green lines, respectively. The dashed red line represents the pp $\Upsilon$ signal fit normalized at the $\Upsilon(1S)$ mass. The lower panel shows the pull distributions. }
\label{fig:10}
\end{center}
\end{figure}

Even if beyond the scope of this contribution, bottomonium states and in particular the $\Upsilon(1S)$, $\Upsilon(2S)$, $\Upsilon(3S)$, represent another crucial observable in QGP-related studies. Due to their quite different binding energy, their melting should occur, according to the original color screening interpretation, at different values of the temperature reached by the system. Therefore, a ``sequential suppression'' of these resonances should be observed when the temperature increases due e.g., to the increased centrality of the collisions under consideration. Even if such an interpretation requires some care, due to the complexity of the evolution of the collision and to the uncertainty in the formation time of the resonances, the LHC results show that the suppression becomes increasingly stronger when the binding energy of the state becomes smaller. This is illustrated in Fig.~\ref{fig:10}, where on the left the centrality dependence of the $R_{\rm AA}$ for $\Upsilon(1S)$ and $\Upsilon(2S)$, measured by CMS, is shown for $\sqrt{s_{\rm NN}}=2.76$ TeV \mbox{Pb-Pb} collisions~\cite{Khachatryan:2016xxp}. In the right panel the dimuon invariant mass spectrum for $\sqrt{s_{\rm NN}}=5.02$ TeV collision is shown~\cite{CMS:2016ayg}, together with the expectations in case of no suppression of the $\Upsilon$ states. The observed suppression is similar to the one observed at the lower LHC energy. 
Remarkably, the weakly bound $\Upsilon(3S)$ states appears to be completely dissolved.

\section{Conclusions}

A wealth of results on charmonium production in nuclear collisions exists today. From the first studies at the SPS to the more recent ones at the LHC, a very rich phenomenology has been experimentally discovered and has represented an excellent ground for the development of sophisticated theoretical approaches that connect the observations with the properties of the QGP. As of today, charmonium production is understood as a balance of suppression effects in the hot medium and (re)generation processes occurring during the deconfined phase and at freeze-out. While all the theoretical calculations require the presence of a long-lived partonic phase in order to be able to reproduce the results, it is fair to say that a quantitative information on the temperature of the system remain difficult to obtain. This is due mainly on one hand to the difficulties in pinning down the melting temperature of quarkonia via lattice calculations (not described in this contribution) and, on the other hand, to the uncertainty
 in the quantitative evaluation of the (re)generation, dominated by the relatively large uncertainties on the open charm cross section evaluation, which is a crucial input to the models. Hopefully, improvements on these specific points in the next years, together with the availability of more accurate data on various observables, and in particular on the charmonium  $v_{\rm 2}$ and on multi-differential suppression patterns, will bring a better understanding of the QGP phase produced in heavy-ion collisions.

\end{document}